\documentstyle[12pt,epsf]{article}
\newcommand{\II}{{\cal I}}
\newcommand{\BB}{{\cal B}}

\newcommand{\NN}{{\cal N}}

\newcommand{\wt}{\widetilde}

\newcommand{\be}{\begin{equation}}
\newcommand{\ee}{\end{equation}}
\newcommand{\ben}{\begin{eqnarray}\displaystyle}
\newcommand{\een}{\end{eqnarray}}
\newcommand{\refb}[1]{(\ref{#1})}

\newcommand{\sectiono}[1]{\section{#1}\setcounter{equation}{0}}

\begin{document}

{}~ \hfill\vbox{\hbox{hep-th/9803194}\hbox{MRI-PHY/P980338}
}\break

\vskip 3.5cm

\centerline{\large \bf Stable Non-BPS States in String Theory}

\vspace*{6.0ex}

\centerline{\large \rm Ashoke Sen
\footnote{E-mail: sen@mri.ernet.in}}

\vspace*{1.5ex}

\centerline{\large \it Mehta Research Institute of Mathematics}
 \centerline{\large \it and Mathematical Physics}

\centerline{\large \it  Chhatnag Road, Jhoosi,
Allahabad 211019, INDIA}

\vspace*{4.5ex}

\centerline {\bf Abstract}

Many string theories contain states which are not BPS, but are
stable due to charge conservation. 
In many cases the description of these
states in the strong coupling limit remains unknown despite
the existence of a weakly coupled dual theory. However, we
show that in some cases
duality symmetries in string theory do enable us to identify
these states in the strong coupling limit and calculate their
masses. We also speculate that in some of the
other cases the missing states
might arise from non-supersymmetric analog of D0-branes.

\vfill \eject

\tableofcontents

\baselineskip=18pt

\sectiono{Introduction} \label{s1}

During the last few years significant progress has been made
towards
understanding the strong coupling dynamics of string theories
using duality symmetries which relate a strongly coupled string
theory to a weakly coupled string/supergravity theory. However,
despite this progress, some of the simple questions concerning
the strongly coupled string theory remain unanswered in many
cases. One such question will be the topic of discussion of this
paper.

Spectra of string theories often contain a special class of
states 
known as BPS states\cite{WITOLI}. 
The mass of a BPS state is completely determined by its
charge. Furthermore, BPS states are stable, and
the degeneracy of a BPS state is independent of all
continuous parameters of the theory, including the coupling
constant.
Due to these properties, the spectrum of BPS states in a string
theory can, in principle, be reliably calculated at all values of
the coupling
by analyzing the theory at weak coupling.
However, quite often a string theory also contains states which are
stable but not BPS. The stability of these states follows
from simple argument involving charge conservation; they are the
lighest states carrying a given set of charges, and hence cannot
decay. The masses of these states at weak coupling can be
calculated reliably using string perturbation theory. The
question that we shall be interested in is: what happens to 
these states at strong coupling? In this case there is
no analog of the BPS formula that gives a relation between the
mass and the charge at all values of the coupling; and
calculating the masses of these states at strong coupling 
requires an explicit study of the strong
coupling dynamics of the theory. The question of degeneracy of
these states is easier to settle; generically we would expect
that the lightest states
carrying a given charge will belong to a single supermultiplet,
since any accidental degeneracy of these
states at the string tree level will be lifted by quantum
corrections. 

The most familiar example of this kind is seen
in the Spin(32)/$Z_2$
heterotic string theory\cite{HETEROTIC} $-$ the states in the
spinor representation of the gauge group are stable due to charge
conservation but are not BPS states. In the fermionic
formulation of the heterotic string theory these states are
obtained by putting periodic boundary condition on all the
left-moving fermions, and have mass of the order of the string
scale. As we increase the string coupling constant, the masses of
these states will get renormalized, but the lightest states
transforming in the spinor representation of Spin(32)/$Z_2$ will
remain stable. Let us denote the mass of such a state by:
\be \label{e1}
(\alpha')^{-1/2} f(g_S)\, ,
\ee
where $(2\pi\alpha')^{-1}$ is the string tension, $g_S$ is the
string couping constant, and $f(g_S)$ is some function of $g_S$
that can be calculated in string perturbation theory. 
The question we are interested in is: how does $f(g_S)$ behave
for large $g_S$?

Since for large $g_S$ this theory can be described by weakly
coupled type I string theory\cite{WITTEND}, one might hope that
the answer can be found by analysing this weakly coupled 
theory. Unfortunately our understanding of the
solitonic spectrum of weakly coupled type I string theory is not
sufficiently complete so as to enable us to answer this question.
Clearly the states we are looking for
should correspond to a non-supersymmetric soliton of type I
theory belonging to the spinor representation of the gauge group.
At a qualitative level one might even describe these as arising
from small loops of D-strings\cite{DABH,HULL,POLWIT}.
However such a description does not enable us to calculate the 
precise mass of these states. In section \ref{s6} we speculate on
the possible origin of these states from non-supersymmetric
D0-branes in type I string theory.

There are however examples in string theory where strong-weak
coupling duality does help us determine the spectrum of stable
non-BPS states in the strong coupling limit.
These examples involve a Dirichlet
$p$-brane (D$p$-brane)\cite{DBRANE} of type II string theory on 
top of an orientifold
$p$-plane (O$p$-plane)\cite{ORIENT,GIMPOL} with SO projection. 
The world-volume theory of this configuration has an SO(2) gauge
field originating from open strings with both ends lying on the
D$p$-brane. Open string states which originate on the
D$p$-brane, goes once around the O$p$-plane, and ends on the
D$p$-brane are charged under this SO(2) gauge group. We shall
normalize the gauge fields so that this state carries unit SO(2)
charge. When the
D$p$-brane coincides with the O$p$-plane, the classical mass of
this state vanishes. However, the supersymmetric ground state of
this system is projected out by the orientifolding 
operation\cite{GIMPOL}, and
hence in the weak coupling limit the lowest mass state in this
sector has mass of the order of the string scale coming from the
string oscillator contribution. Although this state is not a BPS
state, it is stable under quantum corrections, being the lowest
mass state carrying charge under the SO(2) gauge group. Let us
denote the quantum corrected mass of the lightest charged state
in the world-volume theory by 
\be \label{e2}
(\alpha')^{-1/2} f_p(g_S)\, .
\ee
$f_p(g_S)$ is calculable in string perturbation theory, but we
shall be interested in knowing the behaviour of $f_p(g_S)$ for
large $g_S$.\footnote{Although in this case we are analyzing
a state in the world-volume theory of a brane rather
than in the string theory itself, we can easily regarded this
as a state in a compactified string theory by taking the
directions
transverse to the brane / plane to be compact but large, with other 
D$p$-branes and O$p$-planes situated sufficiently far away.} We
shall show that at least in two cases, $p=6$ and $p=7$, 
$f_p(g_S)$ can be completely determined in the strong coupling
limit with the
help of duality symmetries, although for $p=7$ the meaning of the
strong coupling limit becomes somewhat subtle. 
For $p=4$ the behaviour of $f_p(g_S)$
for large $g_S$ can be determined up to an overall multiplicative
numerical factor. Finally for $p=3$ and $p=5$, the D$p$-brane $-$
O$p$-plane configuration in the strong coupling limit 
can be mapped to appropriate configurations of orbifold /
orientifold planes and D-branes in the dual weakly coupled type IIB
string theory. Unfortunately, the precise identification of the
lighest states carrying SO(2) charge is not known in these dual
descriptions. Thus the situation in these two cases is analogous
to the situation with spinor states in type I string theory. In
section \ref{s6} we speculate on the possible identification of
these states.\footnote{Extremal but non-BPS black holes of the
kind analyzed in ref.\cite{DAMARA} might provide other examples
of this kind in some specific region of the moduli space.}

\sectiono{Dirichlet 6-brane on Top of an Orientifold 6-plane}
\label{s2}

This case was discussed earlier in
\cite{STRONG}, but we include it here for completeness. The
theory under consideration is type IIA string theory. The
orientifold 6-plane is given by:
\be \label{e3}
R^{6,1}\times (R^3/(-1)^{F_L}\cdot \Omega\cdot \II_3)\, ,
\ee
where $R^{6,1}$ is the $(6+1)$ dimensional Minkowski space, $R^3$
is the three dimensional Euclidean space, 
$\II_3$ denotes a
transformation that reverses the sign of all the coordinates of
$R^3$, 
${F_L}$ denotes the
contribution to the space-time fermion number from the
left-moving sector of the world-sheet and $\Omega$ denotes the
world-sheet parity transformation. 
In the strong coupling limit the type IIA string theory
is described by M-theory / eleven dimensional supergravity theory
compactified on a circle of large radius $R$, and the coincident
D6-brane $-$ O6-plane system is described by the double cover
$\bar\NN$ of the Atiyah-Hitchin
space\cite{AH,GIBMAN} in M-theory\cite{SEIWITTH,ENH,STRONG}. The eleven
dimensional metric $g^{(11)}_{MN}$ takes the form:
\be \label{e4}
ds_{11}^2 \equiv \sum_{M,N=0}^{10}g^{(11)}_{MN} dx^M dx^N
= \sum_{\mu,\nu=0}^6 \eta_{\mu\nu}dx^\mu dx^\nu + {R^2\over 4}
ds_{AH}^2\, ,
\ee
where $\eta_{\mu\nu}$ is the (6+1) dimensional Minkowski metric,
and
\be\label{e5}
ds_{AH}^2 = f(\rho)^2 d\rho^2 +  a(\rho)^2 \sigma_1^2
+ b(\rho)^2 \sigma_2^2 + c(\rho)^2 \sigma_3^2\, ,
\ee
where,
\ben \label{e6}
\sigma_1 &=& -\sin\psi d\theta + \cos\psi \sin\theta d\phi\, ,
\nonumber \\
\sigma_2 &=& \cos\psi d\theta + \sin\psi \sin\theta d\phi\, ,
\nonumber \\
\sigma_3 &=& d\psi + \cos\theta d\phi\, ,
\een
$x^0, \ldots x^7, \rho, \theta, \phi, \psi$ denote the
coordinates of the eleven dimensional space, and
$f$, $a$, $b$ and
$c$ are known functions of $\rho$. The range of various
coordinates are as follows:
\ben \label{e7}
&& -\infty < x^\mu < \infty, \qquad \hbox{for} \qquad 
0\le \mu\le 6, \nonumber \\
&& \pi \le \rho < \infty, \qquad 0\le\theta \le \pi, \qquad 0\le
\phi\le 2\pi, \qquad 0\le \psi\le 2\pi\, .
\een
The coordinates $\phi$ and $\psi$ are periodic with period
$2\pi$. Finally, there is an identification of coordinates under
a $Z_2$ transformation $I$ given by:
\be \label{e8}
I: \qquad (\theta, \phi,\psi)\to (\pi-\theta, \pi+\phi, -\psi)\,
,
\ee
with all other coordinates remaining fixed.

Near $\rho=\pi$, the functions $f$, $a$, $b$ and $c$ behave as
\be \label{e9}
f\simeq -1, \qquad a\simeq 2(\rho-\pi), \qquad b\simeq\pi, \qquad
c\simeq -\pi\, .
\ee
Since $a$ vanishes at $\rho=\pi$, the metric appears to be
singular there. However if we introduce new angular coordinates
$\wt\theta$, $\wt\phi$ and $\wt\psi$ through the relations
\ben \label{e10}
\sigma_1 &=& d\wt\psi + \cos\wt\theta d\wt\phi\, ,
\nonumber \\
\sigma_2 &=& -\sin\wt\psi d\wt\theta + \cos\wt\psi \sin\wt\theta d
\wt\phi\, ,
\nonumber \\
\sigma_3 &=& \cos\wt\psi d\wt\theta + \sin\wt\psi \sin
\wt\theta d\wt\phi\, ,
\een
and define $\wt\rho=\rho-\pi$, then near $\rho=\pi$ the Atiyah-Hitchin
metric in this coordinate system takes the form:
\be \label{e11}
ds_{AH}^2 \simeq d\wt\rho^2 + 4\wt\rho^2 (d\wt\psi+\cos\wt\theta
d\wt\phi)^2 + \pi^2 (d\wt\theta^2 + \sin^2\wt\theta d\wt\phi^2)\,
.
\ee
Furthermore the transformation $I$ acts on the new coordinates as
\be \label{e12}
I:\qquad (\wt\theta,\wt\phi,\wt\psi)\to
(\wt\theta,\wt\phi,\wt\psi+\pi)\, .
\ee
Thus identification under $I$ makes $\wt\psi$ an angular variable
with period $\pi$. From \refb{e11} we now see that for fixed
$(\wt\theta,\wt\phi)$, $(\wt\rho,2\wt\psi)$ describes a plane near
the origin in polar coordinates, and for fixed
$(\wt\rho,\wt\psi)$, $(\wt\theta,\wt\phi)$ describes a sphere of
radius $\pi$. Thus near $\wt\rho=0$ the space is non-singular,
and locally has the structure or $R^2\times S^2$. The sphere
$S^2$ parametrized by $(\wt\theta,\wt\phi)$ is a non-trivial two
cycle (of minimal area)
in this space, and is known as the `Bolt'. We shall denote
this by $C_0$. In the metric \refb{e4} this sphere has radius
$\pi R/2$ and hence has area:
\be \label{e13}
A =\pi^3 R^2\, ,
\ee
measured in the eleven dimensional metric.

The space $\bar\NN$ spanned by $\rho$, $\theta$, $\phi$ and
$\psi$ has a self-dual harmonic two form
$\omega_0$ given by\cite{MANSCH,GIBRUB,SDUAL}
\be \label{e14}
\omega_0 = K_0 \exp\Big(-\int_\pi^\rho d\rho' {f(\rho')a(\rho')\over
b(\rho') c(\rho')}\Big) \Big( d\sigma_1 -{f(\rho)a(\rho)\over
b(\rho)c(\rho)} d\rho\wedge \sigma_1\Big)\, ,
\ee
where $K_0$ is a normalization constant. It is natural to choose
$K_0$ such that $\int_{C_0}\omega_0=1$.

Given this description of the O6-plane $-$ D6-brane system in
M-theory, we now try to address the question we had posed. First
we need to understand the origin of the SO(2) gauge field in this
system. This is found as follows. If $C$ denotes the three
form gauge field of M-theory, then we can decompose this as
\be \label{e15}
C = \omega_0(y)\wedge A(x) + \ldots 
\ee
where $y$ denotes the coordinate on the Atiyah-Hitchin space, $x$
denotes the coordinates on $R^{6,1}$, $A$ is a one form field on
$R^{6,1}$ and $\ldots$ denotes an infinite number of other terms
which are not relevant for the present analysis. $A(x)$ is the
world-volume SO(2) gauge field that we have been looking for.

The next question will be: how do we get states charged under
$A_\mu$? Since a membrane is charged under the three form field
$C$, a membrane wrapped around the cycle $C_0$ will carry an
$A_\mu$ charge proportional to
$\int_{C_0}\omega_0$.
Thus this is the M-theory description of the open string state
that carries charge under the SO(2) gauge field. Although our
analysis does not show directly that it carries the right amount
of SO(2) charge, it is clear that this is the state carrying
minimal SO(2) charge in the M-theory description, and hence must
have the right charge. The mass of this state is given by the
product of the membrane tension $T_M$ and the area of the two cycle
$C_0$:
\be \label{e16}
m \simeq T_M A = T_M \pi^3 R^2\, .
\ee 
Note that this mass formula 
will get modified by quantum fluctuations of
the membrane. However, as long as $R$ is large,
the classical contribution \refb{e16} to the mass will be the
dominant contribution, and \refb{e16} will be reliable. There
will also be corrections to the mass formula due to membrane
self-interaction since we have a curved membrane instead of a
planar membrane. However, as long as $R$ is large, the
contribution due to such interactions will also be small.

Let us briefly set $\alpha'=1$ and work in the normalization
convention of \cite{SREVIEW}. In this convention,
the membrane tension measured in the eleven dimensional metric is
given by\footnote{This can be seen by noting that upon
compactification on a circle of radius $R$, the string tension
measured in the eleven dimensional metric is given by $2\pi R
T_M$. Using the relation between the eleven dimensional
metric and the ten dimensional string metric we get the string
tension measured in ten dimensional string metric to be $2\pi
T_M$, which must be equal to $(2\pi)^{-1}$ for $\alpha'=1$.}
\be \label{e17a}
T_M={1\over (2\pi)^2}\, .
\ee
On the other hand, $R$ measured in the eleven
dimensional metric is related to the type IIA coupling constant
through the relation\cite{WITTEND}:
\be \label{e17}
R = g_S^{2/3}\, .
\ee
Finally the mass measured in the type IIA string metric is
obtained from that measured in the eleven dimensional metric by
multiplying the latter by a factor of 
$R^{-1/2}=g_S^{-1/3}$\cite{WITTEND}.  Thus
the mass of the state, measured in the type IIA string metric, is
given by $(\pi g_S/4)$. Finally, restoring the factor of
$\alpha'$ by dimensional analysis, we arrive at the final answer
for the mass of the charged open string state in the strong
coupling limit of type IIA string theory:
\be \label{e18}
m \simeq {\pi\over 4} (\alpha')^{-1/2} g_S
\ee
Thus for $p=6$, $f_p(g_S)$ defined in eq.\refb{e2} is given by
$(\pi g_S/4)$ for large $g_S$.

\sectiono{Dirichlet 7-brane on Top of an Orientifold 7-plane}
\label{s3}

In this case the configuration under study
is a single D7-brane on top of a single O7-plane in type IIB
string theory. We shall use the normalization convention of
ref.\cite{SREVIEW} and set $\alpha'=1$. 
Let us denote by $\lambda$ the complex
scalar field 
\be \label{e21}
\lambda = a + i e^{-\phi} \equiv \lambda_1 + i\lambda_2\, ,
\ee
where $a$ is the RR sector scalar field and $\phi$ is the
dilaton. The string coupling constant $g_S$ is related to the 
imaginary part of $\lambda$ via the relation:
\be \label{e21a}
g_S = (\lambda_2)^{-1}\, .
\ee
The O7 plane in type IIB string theory can be described as the
quotient space:
\be \label{e19}
R^{7,1} \times (R^2/(-1)^{F_L}\cdot\Omega\cdot \II_2)\, .
\ee
Since the space transverse to the O7-plane is two
dimensional, we can label this space by a single complex
coordinate $z$. Let us also denote by $w$ the coordinate on the
covering space $R^2$.
Thus $w$ is related to $z$ by the relation
$z = w^2$.

We shall be interested in studying a
configuration of an O7-plane and a D7-brane situated at $z=0$.
In the weak coupling limit the background around
the D7-brane $-$ O7-plane system is given by
\be \label{e22}
\lambda = -{3\over 2\pi i} \ln (z/L^2) = -{3\over \pi i} \ln
(w/L)\, ,
\ee
\ben \label{e23}
ds_{can}^2 \equiv \sum_{M,N=0}^{9}g^{(can)}_{MN}dx^M dx^N
&=& 
\sum_{\mu,\nu=0}^7 \eta_{\mu\nu} dx^\mu dx^\nu  +
\lambda_2 dw d\bar w \nonumber \\
&=& \sum_{\mu,\nu=0}^7 \eta_{\mu\nu} dx^\mu dx^\nu  
+ {1\over 4} \lambda_2 |z|^{-1} dz d\bar z \, ,
\een
where $L$ is a constant, and $g^{(can)}_{MN}$ denotes the ten
dimensional canonical metric of the type IIB string theory. 
The factor of $-3$ in the expression for $\lambda$ reflects the fact
that the O7-D7 system carries a total of $-3$ units of RR charge,
of which 1 unit comes from the D7-brane and $-4$ units come
from the O7-plane. 
Note that since $\lambda$ does not go to a constant
asymptotically, one cannot use the asymptotic string coupling 
constant as a parameter. Instead we can use $L$ as the parameter
measuring the strength of the string coupling constant. As seen
from \refb{e22}, at a large but fixed value of $z$, $\lambda_2$
decreases as $L$ increases. Thus large
$L$ will correspond to strong coupling limit.
Alternatively, one can `regularize' the asymptotic behaviour of
$\lambda$ by placing three D7-branes at $z=z_0(\equiv w_0^2)$ for
some very large $|z_0|$. Now the background $\lambda$ in the weak
coupling region is given by,
\be \label{ex1}
\lambda= -{3\over 2\pi i} \ln{w^2\over w^2-w_0^2} + i g_S^{-1}\,
,
\ee
where $g_S$ is the asymptotic coupling constant. Thus for
$|w|<<|w_0|$, but sufficiently far away from the origin so that
$\lambda_2$ is large, we get
\be \label{ex2}
\lambda \simeq -{3\over 2\pi i} \ln {w^2\over (-w_0^2)} + i
g_S^{-1}\, .
\ee
\refb{ex2} has the same form as \refb{e22} if we choose
\be \label{e27}
L = i w_0 \exp\Big(-{\pi\over 3 g_S}\Big)\, .
\ee
We can now take $g_S$ as our independent parameter instead of
$L$. However, since \refb{ex1}, \refb{ex2} are valid only when
$\lambda_2$ is large, we need to keep $g_S$ small. The theory is
made strongly coupled by taking $|w_0|$ large.

Since the O7-plane and the D7-brane are coincident, the classical
mass of an open string starting on the D7-brane, going around the
O7-plane and ending on the D7-brane vanishes. However, once
non-perturbative effects are taken into account, the coincident
O7-plane $-$ D7-brane system is replaced by a system of three
mutually non-local seven branes\cite{FTHEORY,BADOSE,DASMUK}, the
location and type of these seven branes being given by the
Seiberg-Witten solution for N=2 supersymmetric SU(2) gauge theory
with one massless hypermultiplet in the fundamental
representation. The open
string state under study is represented by an appropriate
network\cite{NETWORK} of open strings 
stretched between these seven
branes\cite{SBPS,JOHAN,FAYY,ZWIE,KISAS,BERFAY,IMAM,SETH}. The mass of
this network can be found by adding up the masses of each
segment of this network. In the $w$
coordinate system the seven branes are located in the region
where $\lambda_2\sim 1$, {\it i.e.} at $|w|\sim |L|$.
Since in this region $\lambda$ is of
order unity, $g^{(can)}_{w\bar w}$ as well as
the tension of any string of type $(p,q)$
is also of order unity. Thus the open string network has size and
mass of
order $|L|$ measured in the canonical metric. This gives a mass
formula of the form:
\be \label{e28}
m\simeq C|L|\, ,
\ee
where $C$ is a numerical constant. 
This mass formula gets corrected by quantum
fluctuations 
on the stretched string network, and also due to the
interaction between different segments of the network. 
However, as long as $|L|$ is
large, the classical contribution to the mass of the stretched
string network dominates over these corrections, and hence
\refb{e28} remains valid.

Computation of the precise value 
of $C$ requires identifying the open string network that
represents the lightest state carrying SO(2) charge.
This has been carried out in the appendix. The answer is
\be \label{e28a}
C = {\Gamma({1\over 3})\over 4\pi \big(\Gamma({7\over
6})\big)^2} 2^{2/3} 3^{1/2}\, .
\ee

Using eq.\refb{e27} we can express \refb{e28} as
\be \label{e29}
m \simeq C |w_0| \exp\Big(-{\pi\over 3 g_S}\Big)\, .
\ee
In order to calculate $f_p(g_S)$ defined in \refb{e2} for $p=7$,
we need to express this in string metric. In ten
dimensional type IIB string theory the string metric is obtained
from the canonical metric by multiplying the latter by a factor
of $g_S^{1/2}$.  
This gives the following expression for the
mass of the charged open string state on the D7-O7 world-volume:
\be \label{e30}
m \simeq C (\alpha')^{-1/2} \Big|{w_0\over \sqrt{\alpha'}}\Big|
\big(g_S\big)^{-1/4} \exp\Big(-{\pi\over 3 g_S}\Big)\, .
\ee
Note that we have restored the factors of $\alpha'$.
Comparison with \refb{e2} gives:
\be \label{e31}
f_7 \simeq
C \Big|{w_0\over \sqrt{\alpha'}}\Big|
\big(g_S\big)^{-1/4} \exp\Big(-{\pi\over 3 g_S}\Big)\, .
\ee

\sectiono{Dirichlet 4-brane on Top of an Orientifold 4-plane}
\label{s4}

In this case the theory is type IIA string theory, and the
orientifold 4-plane is the space:
\be \label{e33}
R^{4,1}\times (R^5/\Omega\cdot \II_5)\, .
\ee
The strong coupling limit of this theory is M-theory compactified 
on a circle $S^1$ of
large radius $R$. In this limit the Dirichlet 4-brane is described by
a five-brane of M-theory with one direction tangential to the
circle. The orientifold 4-plane, on the other hand, is described
by the space:
\be \label{e34}
R^{4,1}\times S^1 \times (R^5/\sigma\cdot \II_5)\, ,
\ee
where $\sigma$ denotes the transformation that changes the sign
of the anti-symmetric tensor field of M-theory.
Various properties of this orbifold were analyzed in
refs.\cite{DASMUKKT,WITTENKT}.

On the world-volume of the five brane there is a massless rank
two anti-symmetric tensor field $\BB_{\mu\nu}$. 
Upon compactification on $S^1$,
this gives the gauge field living on the 4-brane world-volume.
Thus the states charged under this gauge field can be constructed
by taking a string on the five-brane world-volume carrying
$\BB_{\mu\nu}$ charge, and wrapping it on $S^1$. This string in
turn may be constructed as follows. First consider the case where
the five-brane is away from the orbifold fixed plane but parallel
to it. Since membranes can end on five branes\cite{STROM,TOWN}, 
we can consider a membrane configuration
that starts on the five brane, goes around the orbifold
plane, and ends on the five-brane.\footnote{In the covering space
this corresponds to a membrane with one end terminating on the
five-brane and the other end terminating on the image of the five
brane under the $Z_2$ transformation $\sigma\cdot\II_5$.}
This describes a string like
state on the five-brane world-volume carrying $\BB_{\mu\nu}$
charge, with classical contribution
to the tension proportional to
the distance between the five-brane and the orbifold plane. If we
now consider the limit when the distance between the five-brane
and the orbifold plane goes to zero, the classical contribution to the
tension goes to zero, but there is a quantum contribution which,
on purely dimensional grounds, must be of order $m_p^2$, where
$m_p$ is the eleven dimensional Planck mass. This will describe a
non-BPS string on the world-volume of the 5-brane $-$
orbifold plane system.

As in section \ref{s2}, let us briefly set $\alpha'=1$. As argued
in that section, this would correspond to setting the membrane
tension $T_M$, measured in the eleven dimensional metric, to
$1/4\pi^2$.
Since $T_M\propto m_p^3$, this would also correspond to setting
$m_p$, measured in the eleven dimensional metric, to a number of
order unity. Thus in this metric the tension of the non-BPS string
carrying $\BB_{\mu\nu}$ charge is of order unity, and the mass of
the state obtained by wrapping this string around a circle of
radius $R$ is of order $R$. 
Using eq.\refb{e17} we see that this gives a mass of order
$g_S^{2/3}$. In order to convert it to the mass measured in
string metric, we need to multiply it by a factor of $g_S^{-1/3}$.
This gives the follwing expression for the mass of the lightest
state charged under the SO(2) gauge field on the D4-O4 brane
world volume:
\be \label{e36}
m\simeq K (\alpha')^{-1/2} g_S^{1/3}\, ,
\ee
where $K$ is a numerical constant. Note that we have restored the
factor of $\alpha'$ in \refb{e36}. Determining the value of $K$ will
require computing the tension of the non-BPS string through 
a detailed analysis of the dynamics of the membrane
stretched between the five-brane and the orbifold plane. This is
beyond the scope of the present analysis.

Comparing \refb{e36} with \refb{e2} we arrive at the following
result for the strong coupling behaviour of $f_4(g_S)$:
\be \label{e37}
f_4(g_S)\simeq K g_S^{1/3}\, .
\ee

\sectiono{Other D$p$-brane $-$ O$p$-plane Systems} \label{s5}

For $p=5$,
the theory under consideration
is type IIB string theory, and the orientifold 5-plane is
given by the quotient:
\be \label{e32}
R^{5,1} \times (R^4/\II_4\cdot\Omega)\, .
\ee
In studying the strong coupling limit of the coincident O5-D5
system, we can use the S-duality transformation of type IIB
string theory. This converts the transformation $\Omega$ to
$(-1)^{F_L}$\cite{DUORBI}. Thus the strong coupling dual of the
O5-plane is
\be \label{e32a}
R^{5,1} \times (R^4/\II_4\cdot (-1)^{F_L})\, .
\ee
It turns out that this orbifold actually represents the dual of
an O5-plane $-$ D5-brane system\cite{KUTASOV,DUORBI} and not of an
isolated O5-plane. To see
this note that the twisted sector states of this orbifold give
rise to a U(1) gauge field living on the fixed plane. Since an
isolated orientifold plane does not have any massless field living 
on its 
world-volume, \refb{e32} must be dual to an O5-plane $-$ D5-brane
system.
Indeed, the full spectrum of massless twisted sector states in
the orbifold \refb{e32a} consists of
a matter multiplet of the (1,1) supersymmetry algebra in
six dimensions, exactly matching the spectrum of massless states
on the world-volume
of the O5-D5 system.

Once we have identified the strong coupling dual of the O5-brane
$-$ D5-plane configuration, we can formulate the problem under
study in a precise manner. We need to find the lightest state in
this orbifold theory which is charged under the gauge field
coming from the twisted sector of the orbifold. Unfortunately,
even though the question is well posed, the answer is not known.
We shall speculate on a possible answer in the next section.

For $p=3$,
the theory under consideration
is again type IIB string theory, and the orientifold 3-plane is
given by the quotient:
\be \label{e32b}
R^{3,1} \times (R^6/\II_6\cdot\Omega \cdot (-1)^{F_L})\, .
\ee
In studying the strong coupling limit of the coincident O3-D3
system, we can use the S-duality transformation of type IIB
string theory. This leaves invariant
$\Omega\cdot(-1)^{F_L}$, and also the D3-brane.
Thus the D3-O3 system is self-dual. However, this duality
transformation induces an electric-magnetic duality
transformation on the world-volume theory\cite{GREGUT,TOWN}, so that
the state electrically charged under SO(2) gets mapped to a state
magnetically charged under SO(2). Thus the problem of
determining the mass of the lightest electrically charged state
in the strong coupling limit gets mapped to the problem of
finding the mass of the lightest magnetically charged state on
the world-volume of the D3-O3 system in the weak couping limit.
Again this problem, although well-posed, has not been solved.

For $p<3$ a state carrying SO(2) gauge charge has infinite energy
due to the long range gauge field strength that it produces on
the brane.
Thus the mass of such a state is not well defined. This 
leaves only the
case of $p=8$. In this case the theory under study is type IIA
string theory, and its strong coupling limit is described by
M-theory; but unfortunately neither the D8-brane nor the O8-plane
has been understood well from the point of view of M-theory. Thus
we cannot use the M-theory description of the theory to analyze
the spectrum of stable non-supersymmetric states in the
world-volume of this theory.

\sectiono{Some Speculations about the Missing States} \label{s6}

In this section we shall make some speculations about the origin
of the missing states. First let us recollect that in the
examples discussed in this paper, there are three cases where the
required charged states were not found. They are as follows:
\begin{enumerate}
\item
In the weakly coupled type I string theory, we should have
solitonic states transforming in the spinor representation of the
gauge group. These states have not been constructed explicitly.

\item On the orbifold plane $R^{5,1}\times (R^4/(-1)^{F_L}\cdot
\II_4)$ in type IIB string theory, we should have states carrying
charge under the gauge field originating in the twisted sector of
the orbifold. These states have not been found.

\item Finally, in the system containing a D3-brane on top of an
O3-plane, we expect to find states which are magnetically charged
under the SO(2) gauge field living on the world-volume of the
system. These states have also not been found explicitly.

\end{enumerate}

There are two features common between these three cases.
First of all, in each case the theory under
consideration is an orbifold / orientifold of type IIB string
theory. (We are regarding type I theory as type IIB theory modded
out by the transformation $\Omega$.) Second, in each case,
following the duality transformation that led to the existence of
these states, we see that the states under consideration should
correspond to microscopic closed / open D-strings. In the first
case, the states in the spinor representation in the
original $Spin(32)/Z_2$
heterotic string theory are elementary string states and hence
can be regarded as coming from small loops of fundamental
strings. Thus in the dual type I theory these should be
represented as small loops of D-strings. In the second and the
third case, the states under consideration in the original type IIB
theory come from microscopic open strings with ends lying on the
D-brane. Since the strong-weak coupling duality converts a
fundamental string to a D-string, we would expect that in the
dual type IIB theory these states should be represented as
microscopic open D-strings.

Although we do not know how to analyze microscopic D-strings,
there is one qualitative feature of these states that we can
guess $-$ these states must be able to
support configurations where one end of a fundamental string is
stuck to the state. In other words, these states must have
properties similar to a D0-brane! Of course in type IIB theory
itself D0-branes cannot be defined consistently; however one
might conjecture 
that in the presence of orientifold / orbifold planes we
can define analogs of D0-branes in type IIB theory if the
D0-branes are embedded in the orientifold / orbifold plane. 

There are also other arguments one can give in support of this
conjecture. For example a D0-brane of this kind in type I theory
will support open strings with one end stuck to the 0-brane, and
other end free with Neumann boundary condition. These will
transform in the fundamental representation of SO(32) due to the
Chan-Paton factors associated with the free end, and quantizing
the zero energy fermionic states of this 
string will indeed give states of the D0-brane
in the spinor representation of 
the gauge group. (Of course the origin of the GSO projection in
this case is somewhat obscure, and presumably related to the
existence of a subtle $Z_2$ gauge symmetry on the D0-brane
world-volume\cite{POLWIT}.) In the second
example, a single D0-brane situated on the orbifold 5-plane
will most likely carry charge under the gauge field coming from
the twisted sector as in the case of \cite{DOUMOO}. This will
again be in accordance with the expected property of these 0-branes.

Conformal field theories describing excitations on these
D0-branes are certainly going to be more complicated than those
for supersymmetric D-branes.
Construction of these conformal field theories remains a
challenging problem for the future. 

\noindent {\bf Acknowledgement}: A preliminary version of this
work was reported in the string workshop at the Institute of
Physics, Bhubaneswar. I wish to thank the workshop participants
for their comments.

\appendix

\sectiono{Computation of the Coefficient $C$} \label{a1}

In this appendix we shall compute the coefficient $C$ introduced
in section \ref{s3}. We begin by introducing some notations. A
$(p,q)$ string will denote a bound state of $p$ fundamental
strings and $q$ D-strings, and a $(p,q)$ seven brane will denote
a seven brane on which a $(p,q)$ string can end. Thus in this
notation a D7-brane will be labelled as a (1,0) seven brane. The
SL(2,Z) monodromy around a $(p,q)$ seven brane is given by 
\be \label{ea1}
M_{p,q} = \pmatrix{1-pq & p^2 \cr -q^2 & 1+pq}\, .
\ee
Our convention is the same as that in ref.\cite{ZWIE};
as an $(r,s)$ string / 7-brane crosses the cut associated with a
$(p,q)$ seven brane in the anti-clockwise direction, it gets
converted to an $M_{p,q}\pmatrix{r\cr s}$ string / 7-brane. 
\begin{figure}[!ht]
\begin{center}
\epsfbox{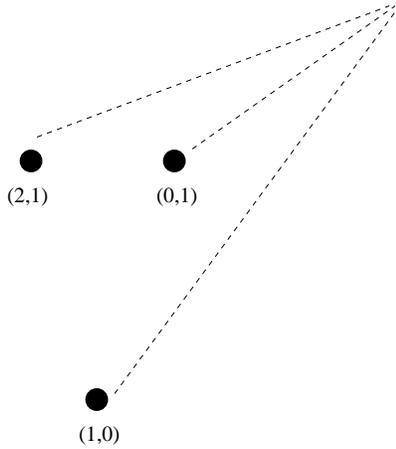}
\end{center}
\caption[]{\small D7-O7 system with D7-brane far away from the
O7-plane. The dashed lines indicate the path along which the
seven branes are viewed.}
\label{f1}
\end{figure}
\begin{figure}[!ht]
\begin{center}
\epsfbox{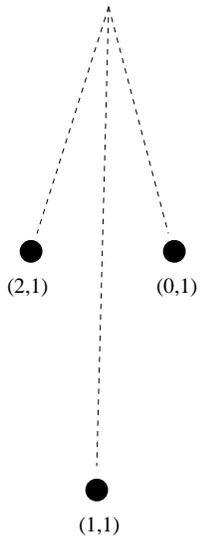}
\end{center}
\caption[]{\small A different view of the configuration shown in
Fig.\ref{f1}.}
\label{f2}
\end{figure}

Let us now turn to the O7-D7 system under study. First consider
the case where the D7-brane is far away from the O7-plane. In
this case the D7-brane is represented as a (1,0) 7-brane, whereas
the O7-plane splits into a pair of parallel (0,1) and (2,1)
7-branes\cite{FTHEORY} as shown in Fig.\ref{f1}.
In order to make this description meaningful, we must describe
the path along which the asymptotic observer views different
7-branes. This has been indicated by the dashed lines in
Fig.\ref{f1}. The same configuration can be viewed as a
collection of a (1,1), (0,1) and (2,1) 7-branes by changing the
path along which we view the branes. This has been shown in
Fig.\ref{f2}.
\begin{figure}[!ht]
\begin{center}
\epsfbox{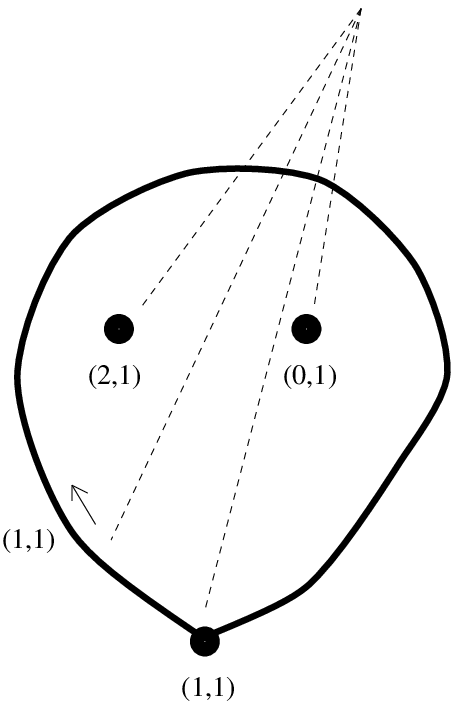}
\end{center}
\caption[]{\small An open string configuration representing a
state carrying SO(2) charge.}
\label{f2a}
\end{figure}
\begin{figure}[!ht]
\begin{center}
\epsfbox{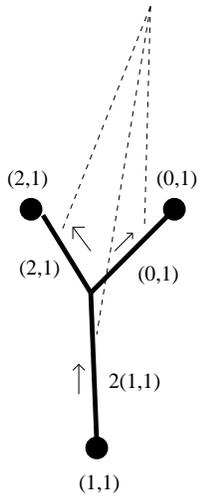}
\end{center}
\caption[]{\small The minimal energy configuration of open string
network representing a state carrying SO(2) charge. Although in
this figure we have displayed the open strings to be lying along
straight lines, they actually lie along geodesics described in
eq.\refb{ea9}.}
\label{f3}
\end{figure}

A state on the D7-O7 world volume theory, carrying charge under
the world-volume gauge field, is represented by an open string
starting on the (1,1) 7-brane,  going
around the (2,1) and the (0,1) 7-branes, and coming back to the
(1,1) brane, as shown in Fig.\ref{f2a}. However, as was pointed
out in ref.\cite{ZWIE}, this configuration does not represent a
geodesic, and hence does not minimize the energy stored in the
string. The minimal energy configuration\cite{BERFAY} is shown in
Fig.\ref{f3}. This configuration can be obtained from the one in
Fig.\ref{f2a} by deforming the open string loop through the (2,1)
and the (0,1) seven branes. According to the rules of
ref.\cite{ZWIE}, during this process new open string prongs
connecting the original string and the 7-branes are generated, as
has been shown in Fig.\ref{f3}. Note that the string junction
where the (2,1), (1,1) and the two (1,1) strings meet conserve
charge, as required. However, in checking charge conservation one
must ensure that all the strings meeting at a junction are viewed
along the same path, or along paths which are continuously
deformable to one another without passing through a seven-brane.
This is satisfied in Fig.\ref{f3}, as the three dashed lines
along which the three types of strings are viewed, are
continuously deformable to one another. The precise location of
the string junction, as well as the path followed by the
different segments of the network is determined by minimizing the
energy of this configuration following refs.\cite{SBPS,BERFAY}.
If we naively follow the arguments of ref.\cite{BERFAY} by
replacing their three brane by a D7-brane, we would conclude
that this
configuration is supersymmetric. However, a more careful
analysis\cite{SETH} shows that it does not represent a BPS state.
This is in accordance with the expected property of this state.

\begin{figure}[!ht]
\begin{center}
\epsfbox{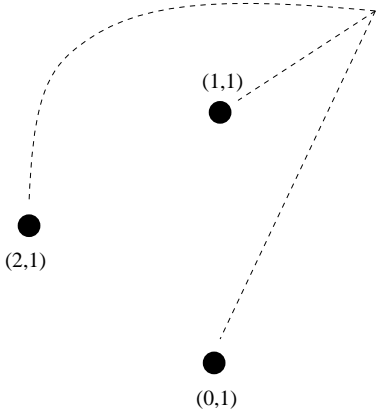}
\end{center}
\caption[]{\small The 7-brane configuration for coincident D7-O7
system.}
\label{f4}
\end{figure}
\begin{figure}[!ht]
\begin{center}
\epsfbox{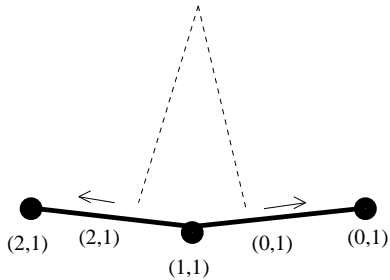}
\end{center}
\caption[]{\small The open string network representing the lightest
charged state at critical distance.}
\label{f5}
\end{figure}
Now consider reducing the distance between the D7-brane and the
O7-plane to zero.\footnote{Since the O7-plane splits into two
branes, one cannot make this distance vanish. What we mean by
this limit is that the parameter that labels the classical
distance between the plane and the brane goes to zero.} 
In this limit the
configuration of 7-branes is given by the Seiberg-Witten result 
for N=2 supersymmetric SU(2) gauge theory with one flavour of
massless quark hypermultiplet in the fundamental
representation\cite{SEIWIT}. This configuration is displayed in
Fig.\ref{f4} and is reached from Fig.\ref{f2} by a continuous
deformation
moving the (1,1) seven brane through the other two seven branes. 
The question that we would like to address is: what configuration
does the string network shown in Fig.\ref{f3} evolve to during
this process? For this we try to follow this configuration as
the distance between the O7-plane and D7-brane decreases
continuously. At some point during this process, which we might
call the critical distance, the length of the (1,1) portion of
the network goes to zero, as shown in Fig.\ref{f5}. The question
is what happens to this network when we reduce the distance even
further?
\begin{figure}[!ht]
\begin{center}
\epsfbox{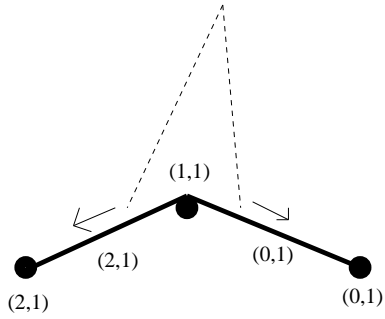}
\end{center}
\caption[]{\small A possible open string network representing
lightest charged state at less than the critical distance.}
\label{f6}
\end{figure}
\begin{figure}[!ht]
\begin{center}
\epsfbox{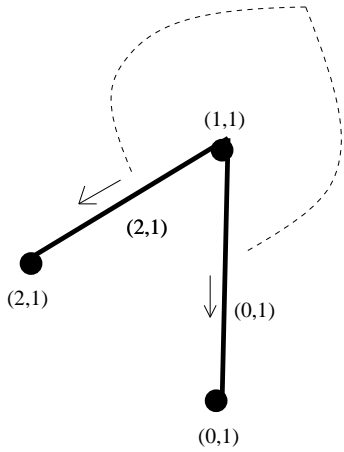}
\caption[]{\small A possible open string network representing
lightest charged state for coincident D7-O7 system.}
\label{f7}
\end{center}
\end{figure}

There are two possibilities we can consider. The first
possibility, which is a conservative picture, is that as the
distance parameter is reduced below the critical distance, the
(2,1) and the (0,1) strings are dragged along with the (1,1)
seven brane as shown in Fig.\ref{f6}. According to this picture,
the lightest charged state on the D7-O7 world volume for
coincident D7-O7 system is given by the open string network shown
in Fig.\ref{f7}. This is certainly an allowed configuration, and
hence the mass of this state gives an upper bound to the mass of
the lightest charged state in the world-volume theory.
\begin{figure}[!ht]
\begin{center}
\epsfbox{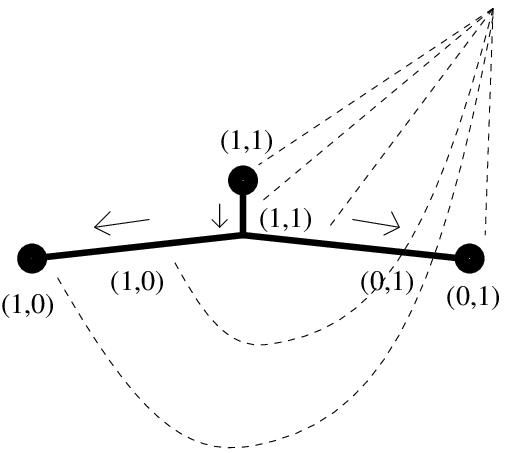}
\caption[]{\small Another possible open string network representing
lightest charged state at less than critical distance.}
\label{f8}
\end{center}
\end{figure}

A more radical possibility for what happens at less than critical
distance has been shown in Fig.\ref{f8}. First of all, note that
in this figure the original (2,1) seven brane and the attached (2,1)
string are viewed along a different path in order to be able to
check charge conservation at the string junction. Thus they
appear as (1,0) seven brane and (1,0) string respectively. The
second point to note is that charge conservation at the string
junction demands that now there is a single (1,1) string coming
out the (1,1) seven brane instead of two (1,1) strings as in
Fig.\ref{f3}. One way to interprete this phenomenon is that the
original
(2,1) string, in the process of passing through the (1,1) seven
brane, has developed a (1,1) prong connecting it to the (1,1)
seven brane. This new prong cancels the charge of one of the
original (1,1) strings, leaving behind only one (1,1) string
joining the (1,1) seven brane to the triple string junction.
Deformations of this kind, leading from Fig.\ref{f3} to
Fig.\ref{f8}, have been discussed recently in \cite{SETH}.
\begin{figure}[!ht]
\begin{center}
\epsfbox{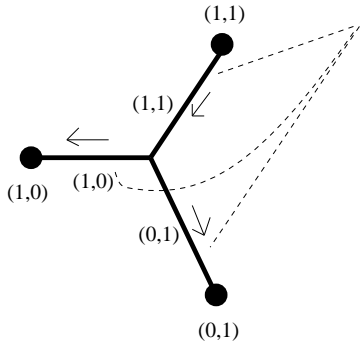}
\caption[]{\small The open string network representing
lightest charged state for coincident D7-O7 system.}
\label{f9}
\end{center}
\end{figure}

Naively the configuration displayed in Fig.\ref{f8} seem to
carry only
half the SO(2) charge as the configuration of Fig.\ref{f3}, since
only one (1,1) string comes out of the (1,1) seven brane.
However, since the configuration of
Fig.\ref{f8} is obtained by continuous deformation from the
configuration of Fig.\ref{f3}, we would expect Fig.\ref{f8} to
represent a state carrying the same SO(2) charge as Fig.\ref{f3}.
Also by embedding the D7-O7 configuration under study into a
complete orientifold of type IIB on $R^{7,1}\times
T^2/(-1)^{F_L}\cdot\Omega\cdot\II_2$, it is easy to see that a
state carrying half integral SO(2) charge (and no other charge)
cannot exist for purely
group theoretic reason. Thus the most likely interpretation of
the state displayed in Fig.\ref{f8} is that it represents a
state carrying unit SO(2) charge. This charge assignment cannot
be directly read out from examining how many (1,1) open
strings end on the (1,1) seven brane. Similar situation was
encountered in ref.\cite{ZWIE} in searching for $E_n$ gauge boson
in the orientifold theory.
According to this view, the lightest open string state carrying
unit SO(2) charge for coincident D7-O7 system is given by the
configuration shown in Fig.\ref{f9}. The $Z_3$ symmetry of the
brane configuration\cite{SEIWIT} implies that the three string
vertex must be located at the center of the equilateral triangle
whose vertices are the locations of the three seven branes. 
The coefficient $C$ is determined by calculating the mass of the
open string network of Fig.\ref{f9}. This is the problem to which
we shall now turn.

The non--perturbative effects modify the background fields
\refb{e22} \refb{e23} to\cite{FTHEORY}
\be \label{ea2}
\lambda(z) = \Big( {da_D\over dz}\Big) / \Big( {da\over
dz}\Big)\, ,
\ee
\be \label{ea3}
ds_{can}^2 = 2\lambda_2 da d\bar a + 
\sum_{\mu,\nu=0}^7 \eta_{\mu\nu} dx^\mu dx^\nu\, ,
\ee
where $a$ and $a_D$ are integrals of Seiberg-Witten
differential\cite{SEIWIT}, appropriately normalized so that the
background described in \refb{ea2}, \refb{ea3} matches the one
given in \refb{e22}, \refb{e23} for large $|z|$. This can be
achieved by choosing
\be \label{ea4}
a(z) = \wt L a^{(1)}(z/\wt L^2), \qquad a_D(z) = \wt L
a_D^{(1)}(z/\wt L^2), 
\ee
where $a^{(1)}$ and $a_D^{(1)}$ are the functions defined
in ref.\cite{BILFER} and $\wt L$ is a constant to be determined
shortly. $a^{(1)}(u)$ and $a_D^{(1)}(u)$ have the following asymptotic
behaviour for large $|u|$:
\ben \label{ea5}
a^{(1)}(u) &\simeq& \sqrt{u\over 2} \, , \nonumber \\
a^{(1)}_D(u) &\simeq & {3i\over 2\pi} \sqrt{u\over 2} \Big[ \ln u
+{4\over 3} \ln 2 -2 + \ln 3 + {i\pi\over 3}\Big]\, .
\een
Using eqs.\refb{ea4}, \refb{ea5} we see that the background
described in \refb{ea2}, \refb{ea3} matches the one given in
\refb{e22}, \refb{e23} for large $|z|$ if we choose:
\be \label{ea6}
\wt L = 2^{2/3} 3^{1/2} e^{i\pi/6}\, L\, .
\ee

In units where $\alpha'$ is set equal to unity, the tension of a
$(p,q)$ string, measured in the canonical metric, is given
by\cite{SCHWARZTEN}
\be \label{ea7}
T(p,q) ={1\over 2\pi} {1\over\sqrt{\lambda_2}}|p + q\lambda|\, .
\ee
Thus the mass of a $(p,q)$ string stretched between $z$ and
$z+dz$ is given by
\be \label{ea8}
T(p,q)\, \sqrt{g_{z\bar z}}\,  |dz| = {1\over \sqrt 2 \pi} 
|p\, da + q\, d a_D|\, .
\ee
As a result, for a $(p,q)$ string stretched between two fixed
points in the $z$ plane, the minimum energy configuration will
require the string to lie along the curve\cite{SBPS}
\be \label{ea9}
(pa + qa_D) = C_1 + C_2 t\, ,
\ee
where $C_1$ and $C_2$ are constants and $t$ is a real parameter
labelling points on the curve. We shall call this the $(p,q)$
geodesic. This gives the following expression for the mass of a
$(p,q)$ string stretched between two points $z_1$ and
$z_2$\cite{SBPS}:
\be \label{ea10}
{1\over \sqrt 2\pi} |pa(z_1) + qa_D(z_1) - pa(z_2) - qa_D(z_2)|
= {|\wt L|\over \sqrt 2\pi} |pa^{(1)}(u_1) + qa^{(1)}_D(u_1
) - pa^{(1)}(u_2) - qa^{(1)}_D(u_2)|\, ,
\ee
where $u=z/\wt L^2$. 
The mass of the open string configuration in Fig.\ref{f9}
can be calculated simply by adding up the contributions
from the different segments of the string network. The
calculation is further simplified by noting that the combination
$(pa+qa_D)$ vanishes at the location of the $(p,q)$ seven brane.

{}From \cite{BILFER} we learn that the (1,0), (0,1) and (1,1)
seven branes in Fig.\ref{f9} are located at $u=-1$, 
$u=e^{-i\pi/3}$ and $u=e^{i\pi/3}$
respectively. Thus due to the $Z_3$ symmetry of this
configuration, the triple string junction 
in this figure is located at $u=0$. Using the same
symmetry we can also conclude that the three 
open string segments in
this diagram have equal mass. Thus the mass of the open string
configuration displayed in Fig.\ref{f9} is given by:
\be \label{ea11}
m={1\over \sqrt 2\pi} |\wt L| |3a_D^{(1)}(0)|\, .
\ee
According to \cite{BILFER} $|a_D^{(1)}(0)|$ is given by
$\sqrt{2}F({5\over 6}, {5\over 6};2;1)/12=\sqrt{2}\Gamma({1\over
3})/12(\Gamma({7\over 6}))^2$.
Using \refb{ea6} we can now rewrite \refb{ea11} as
\be \label{ea12}
m=
{\Gamma({1\over 3})\over 4\pi (\Gamma({7\over 6}))^2} 
\cdot 2^{2/3} \cdot 3^{1/2} \cdot
|L|\, .
\ee
Comparing this with \refb{e28} we get,
\be \label{ea13}
C=
{\Gamma({1\over 3})\over 4\pi (\Gamma({7\over 6}))^2} 
\cdot 2^{2/3} \cdot 3^{1/2}\, .
\ee

Since this derivation of the value of $C$ requires the mild
assumption that the configuration of Fig.\ref{f3} smoothly
evolves into the configuration of Fig.\ref{f9}, we shall now also
derive an absolute upper bound on the value of $C$ by  
calculating the mass of the open string configuration of
Fig.\ref{f7}. Using the symmetries of this configuration,
and eq.\refb{ea10}, we
see that the mass of this open string configuration is given by:
\be \label{ea14}
{1\over \sqrt 2\pi} |\wt L| |2a_D(e^{i\pi/3})|\, .
\ee
$|2a_D(e^{i\pi/3})|$ is  equal to $3\sqrt 2/\pi$\cite{BILFER}.
This gives the mass of the open string configuration of
Fig.\ref{f7} to be
\be \label{ea15}
{1\over \pi^2} \cdot 2^{2/3} \cdot 3^{3/2} \cdot
|L|\, .
\ee
This gives,
\be \label{ea16}
C \le
{1\over \pi^2} \cdot 2^{2/3} \cdot 3^{3/2}\, .
\ee

\end{document}